\documentstyle[aps,epsf,multicol]{revtex} 
\begin{document} 

\title{
      Effect of Vortex Screening on the Bose Glass
      to Entangled Liquid Transition of Flux Lines in Superconductors
      }

\author{Ajay Nandgaonkar$^{1}$, D. G. Kanhere$^{1}$ and 
        Nandini Trivedi$^{2}$} 

\address{$^{1}$ Department of Physics, University of Pune, Pune 411007, India \\
         $^{2}$ Department of Theoretical Physics,
                Tata Institute of Fundamental Research, 
                Mumbai 400005, India 
        } 

\date{\today} 

\maketitle 

\begin{abstract} 
We study the phase diagram of flux lines in superconductors
with columnar pins.
Based on  numerical exact diagonalisation simulations 
on small clusters, we get two phases of vortices:
A low temperature pinned glass with diverging tilt modulus
and a high temperature delocalised entangled vortex liquid.
For random potential disorder 
we find a new phase transition temperature $T_{\rm BG}$ 
from a pinned Bose glass to an entangled liquid 
that reduces with increasing vortex density.
This occurs primarily because vortices screen the disorder 
potential and generate an effective weaker random potential 
with increasing vortex density.
For a fixed fraction of randomly placed attractive columnar
pins of strength $V_{\rm pin}$, we find a Mott insulating 
phase when the vortex density exactly matches the number 
of pins  ($B = B_{\phi}$). 
We also find a transition from
a strongly pinned Bose glass for $B < B_{\phi}$ to a 
weakly pinned Bose glass for $B > B_{\phi}$ as the vortex 
density is varied.

%
%
\end{abstract} 
 
\begin{multicols}{2}  
 
%
%
\section{Introduction} 
\label{sec:intro} 

The interest in equilibrium and dynamical properties of 
flux arrays in type II superconductors 
originates both from their technological importance 
and the rich variety of behavior these systems exhibit under different
physical conditions. 
In a classic paper Abrikosov in 1957 \cite{abrikosov} showed that 
for applied fields $H$ such that $H_{c1} < H < H_{c2}$,  the magnetic
field penetrates in the form of quantized flux tubes ($\phi_0 = hc/2e$), 
which in the absence of disorder form a triangular lattice. 
In conventional low temperature superconductors, this Flux Line Lattice
(FLL) was believed to exist at all temperatures upto $H_{c2}(T)$. 
With the discovery of high ${\rm T_c}$ superconductors, it was
realized that, due to much higher transition temperatures, 
reduced effective dimensionality 
and short coherence lengths, fluctuations play an 
important role in deciding the structure and 
dynamics of FLLs \cite{blatter}. 
One of the most significant consequences of enhanced thermal 
fluctuations on the FLL is its melting into a flux liquid
phase via a first order transition \cite{safar,expt1}. 

Disorder arising from vacancies and interstitials, twin boundaries, 
grain boundaries and columnar pins also modify the structure
and dynamics of the vortex lattice. 
The presence of strong static disorder is technologically 
relevant, for it leads to effective pinning of vortices 
thereby leading to high critical currents \cite{blatter}.
It also can lead to novel glassy phases such as the 
vortex glass and Bragg glass for the case of {\em random} quenched disorder 
\cite{fisher-huse,larkin,giamarchi,gaifullin}.  

Columnar defects i.e. linear damaged tracks in the material caused 
by heavy ion irradiation have emerged as very effective pinning
centers \cite{blatter,civale}. 
In order to model the effect of columnar pins on the FLL, 
we exploit the mapping of 3D interacting flux lines onto
bosons in (2+1)D \cite{nelson1,nelson2}. 
In the mapped quantum problem, the 
columnar defects naturally map onto a time-independent 
random potential for bosons \cite{nelson-vinokur}. 
The irreversibility line can be interpreted as a phase transition 
where the vortex liquid at high temperatures freezes into a 
{\em Bose glass} (BG) of vortices pinned by columnar pins 
at lower temperatures. 
Path integral Monte Carlo simulations \cite{sen} 
find a low temperature BG with 
patches of ordered region with positional {\em and} orientational
order which melts into an entangled defected liquid at
high temperatures. 
Also, when the vortex density and 
defect densities are equal (at the matching field $B_{\phi}$), 
each flux line is attached to one pin, leading to a 
Mott insulator (MI) phase\cite{nelson-vinokur}. 
Such a Mott insulator has been observed in 
magnetization relaxation experiments \cite{kwok1}.

In this paper, we study, using numerical 
exact diagonalisation on small lattices, the different phases
of flux lines with columnar pins for various densities of vortices
and disorder strengths.
We consider a lattice of $N$-sites with 
$N_v = H \times {\rm area} / \phi_0$ vortices, 
interacting via a hard core potential. 
We model disorder in two ways: 
(a) pinning disorder where a finite fraction of attractive pins, 
each of strength $V_{\rm pin}$, are placed randomly; and
(b) a random disorder potential at each point.
In case (a),  we find that an entangled vortex liquid 
is stable against weak pins. 
For high pinning strengths, a Mott insulator is 
realised when the number of vortices
is equal to the number of pins at the matching field $B_{\phi}$.
Signatures of a strongly pinned Bose glass and a weakly pinned
Bose glass are also seen as the vortex density is 
tuned across $B_{\phi}$.
In case (b), we find two phases in the density-disorder plane. 
At low disorder an entangled vortex liquid which localizes into a 
pinned Bose glass with increasing disorder. 
We find that the critical disorder strength required to 
pin the vortex liquid increases with increasing vortex density.
This implies that the temperature required to depin the vortices 
is reduced with increasing fields (see Fig.~\ref{fig:7}). 

We organize the paper in the following way. 
In Section~\ref{sec:model} we give the details of our model.
In Section~\ref{subsec:pin} we discuss our results for 
pinning disorder, where we can access different phases of 
vortices including the Mott insulator. 
In Section~\ref{subsec:box} we discuss our simulations 
for the case where each site has a random disorder potential and 
conjecture an interesting experimental 
implication of our phase diagram. 

%
%
\section{The Model} 
\label{sec:model} 

Consider a system of $N_v$ flux lines in 3D 
in a magnetic field (${\bf B}$) aligned with the $z$-axis,
described by their 2D 
trajectories ${\bf r}_i (z)$ as they traverse a sample of thickness
$L$ with $N_P$ columnar pins. 
Their free energy \cite{nelson-vinokur} is given by 
\begin{eqnarray} 
{\cal F} & = & 
             \int_0^L dz 
             \sum_{i=1}^{N_v} 
             \left\{
                {\tilde{\epsilon}_1 \over 2} 
                \left| {d {\bf r}_i (z) \over dz } \right|^2 
              + {1 \over 2} 
                \sum_{j \ne i}^{N_v} 
                V [ r_{ij}  (z) ]   
                \right. \nonumber \\ 
       &&       \left.
              + \sum_{k=1}^{N_P} 
                V_P [ {\bf r}_i (z) - {\rho}^{\rm pin}_k ]
             \right\}. 
\label{eq:free-energy} 
\end{eqnarray} 
The first term in Eq.~(\ref{eq:free-energy}) is the line tension term 
with tilt modulus $\tilde{\epsilon}_1$. 
The second term denotes the interaction energy of all vortex pairs 
on a constant $z$-plane, where $r_{ij} = \vert {\bf r}_i - {\bf r}_j \vert$
and $V(r)$ the inter-vortex potential.
The last term denotes $N_P$ columnar pins ($\parallel {\bf B}$), 
modeled by $z$-independent potential $V_P$ 
placed on randomly distributed positions $\{ {\rho}^{\rm pin}_k \}$. 

The classical statistical mechanics of Eq.~(\ref{eq:free-energy})
is equivalent to the quantum mechanics of interacting bosons 
interacting with a potential $V(r)$ in 2D with a random static 
potential $V_P ({\bf r})$.
The partition function is determined by the ground-state energy of a 
fictitious quantum hamiltonian \cite{nelson2,nelson-vinokur,nelson}. 
Using this mapping, the thermal fluctuations of the 
3D-vortices get mapped onto the effective quantum fluctuations 
of bosons in two spatial 
dimensions and one imaginary time dimension. 
In this mapping, the temperature of the vortex system $T$ plays the 
role of the Planck number $\hbar$. 
The bending energy $\tilde{\epsilon}_1$ of the flux lines is 
equivalent to the mass $m$ of the bosons, 
so that like a lighter particle, a softer flux-line can wander more. 
The length of the vortices is equivalent to the inverse 
temperature {\em of} bosons so that in order to simulate a thick 
sample $(L \rightarrow \infty)$ one considers the ground state of 
the quantum hamiltonian given by \cite{nelson-vinokur}, 
\begin{equation} 
H =  - t \sum_{ij} (a^{\dagger}_i a_j + h. c.) 
     + {1 \over 2} \sum_{i \ne j} V(r_{ij})
     + \sum_i \mu_i n_i 
\label{eq:Bosehubbard} 
\end{equation} 
where, operator $a_i$ ($a^{\dagger}_i$) annihilates (creates) a boson at 
site $i$, $n_i = a^{\dagger}_i a_i$ is the number operator and $\mu$ is 
the chemical potential at site $i$. 
$t$ is the measure of quantum fluctuations of bosons. 
$\mu_i$ has two parts: a uniform part $\mu \propto (H - H_{c1})$ which 
fixes the flux line density and $\delta \mu_i$ which represents site
disorder. 

The inter-vortex potential can be written as 
$V(r)= 2 \epsilon_0 K_0 (r / \lambda)$, with the modified 
Bessel function $K_0 (x) \propto - \ln(x)$ as $x \rightarrow 0$, 
and $K_0 (x) \propto x^{-1/2} \exp{(-x)}$ for $x \rightarrow \infty$.
The energy scale is set by $\epsilon_0 = (\phi_0 / 4 \pi \lambda)^2$, 
where $\phi_0 = hc/2e$ and $\lambda$ the London penetration depth. 
In the low density limit i.e. when the inter-vortex separation ($a_0$)
is large compared $\lambda$, 
the long range part of $V(r)$ dominates. 
In the absence of disorder, at low temperatures, the vortices form a 
triangular lattice. 
In the high density limit ($a_0 \leq \lambda$), the short ranged
part of $V(r)$ decides the structure and elastic
properties of the flux array. 
In the presence of columnar pins at low density, individual 
vortices get pinned whereas at high density, vortex bundles
are pinned and the system 
breaks up into patches of ordered regions \cite{sen}. 
We use a simplified model and describe the inter boson 
interaction by a hard core interaction. 
In the low density limit, these bosons describe individual
vortices whereas in the high density limit they describe
the renormalized bundles of vortices. 

We fix $t$ to be unity and, disorder $\delta \mu_i$ is introduced in 
two ways: 
(a) {\em Pinning Disorder} : 
Attractive pins of strength $V_{\rm pin}$ 
are introduced at randomly chosen 25\% 
of the sites, and 
(b) {\em Box Disorder} : Random disorder potentials at each lattice 
site chosen either from a uniform distribution of 
width $[- \Delta /2, \Delta/2]$ 
or from a gaussian distribution of width $\Delta$. 
We work on a $4 \times 4$ square lattice with 
periodic boundary conditions, and use exact diagonalization 
techniques (such as Lanczos \cite{dagotto}) to compute the 
ground state properties of (\ref{eq:Bosehubbard}). 

Our aim is to understand the combined effects of the strong
intervortex hard core interaction and columnar disorder on the phases. 
The bosons described by Eq.~(\ref{eq:Bosehubbard}) can exist 
in three different phases \cite{fisher-weichmann}: 
(a) Kinetic energy dominated superfluid phase for bosons that is 
equivalent to a high temperature entangled 
flux liquid in the vortex problem, where flux lines are delocalised and 
can hop freely from one columnar pin to another. 
(b) Disorder dominated Bose glass phase for bosons with a 
vanishing superfluid density $\rho_s$,
that is equivalent to a Bose glass vortex
phase with a diverging tilt modulus  $c_{44}\propto \rho_s^{-1}$.
(c) Interaction dominated gapped Mott insulator phase for 
bosons corresponds to the phase when the vortex density matches the density
of columnar pins in the vortex problem.
As opposed to a BG, the MI is incompressible since the vortex density remains 
locked to the density of pins over a finite range of external fields
\cite{nelson-vinokur} and behaves like a Meissner Phase with 
$B = B_{\phi}$ instead of $B=0$. 

%
%
\section{Results} 
\label{sec:Results}

\subsection{Pinning Disorder} 
\label{subsec:pin} 

\begin{figure}
\epsfxsize=\hsize \centerline{\epsfbox{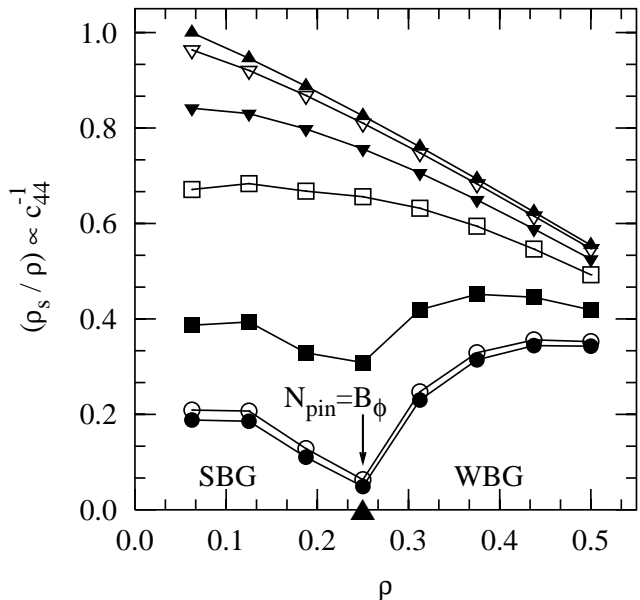}}
\caption{
{\bf Pinning disorder}: 
$\rho_s / \rho$ as a function of $\rho$ for $V_{\rm pin} = 0.2, 
1.0, 2.0, 3.0, 5.0, 7.5, 8.0$ (top to bottom).
{\em Pins} of strength $- V_{\rm pin}$ are placed on randomly
chosen sites. 
$N_{pin} = 4$. 
Data averaged over 300 disorder realizations.
For high $V_{\rm pin}$ the Mott Insulator is realized
at the matching field, when $ \rho = N_{pin}$ (indicated by
filled triangle on the x-axis). 
} 
\label{fig:5} 
\end{figure}

On a $4 \times 4$ square lattice, we randomly choose a fixed fraction 
(say 25 \%, $N_{\rm pin} = 4$) of sites to be pins, each of strength
$- V_{\rm pin}$.  
In Fig.~\ref{fig:5}, we plot the superfluid density $\rho_s / \rho$, 
which is a measure of the stiffness of the phase of the order
parameter, as a function of density $\rho$, for several values 
of  $V_{\rm pin}/t$.  
Since we work on a lattice, $\rho_s / \rho \rightarrow 0$ 
as $\rho \rightarrow 1$, 
and the repulsive local interactions give rise to a Mott insulator
for any values $V_{\rm pin}$. 
In addition, for large $V_{\rm pin}/t$, the superfluid density goes to zero 
at a special Mott insulator point when $N_v = N_{\rm pin} = 4$; 
{\em the matching condition}
where each columnar pin traps exactly one 
vortex. 
The distribution $P(n_i)$, of local density $\langle n_i \rangle$,
follows a bimodal distribution, with a peak at 0 
and another peak at 1. 
Note that the origin of this Mott insulator (for $B = B_{\phi}$) 
is different from the MI obtained in the jamming limit ($\rho \rightarrow 1$).
This Mott insulator can be viewed as a Meissner phase for vortices at the 
matching field instead of zero field \cite{nelson-vinokur,wengel}. 

%
%

\begin{figure}
\epsfxsize=\hsize \centerline{\epsfbox{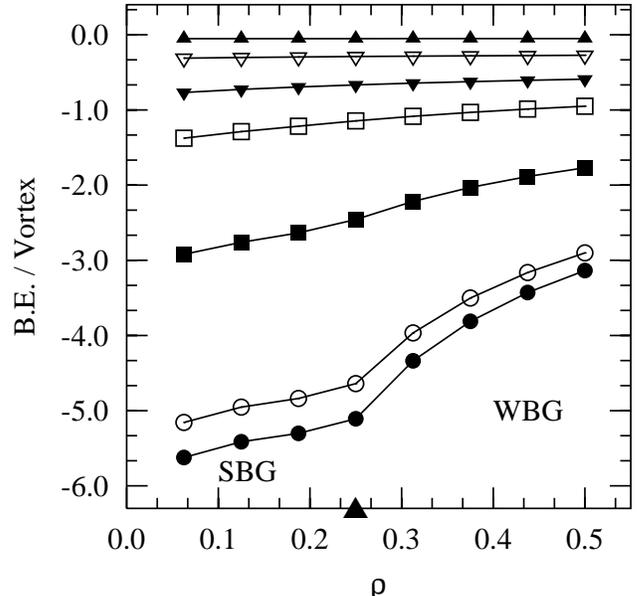}}
\caption{
{\bf Pinning Disorder}: 
Binding energy per vortex as a  
function of density $\rho$ for $N_{pin} = 4$.
$V_{\rm pin}$ values same as in Fig~\ref{fig:5}.
At large $V_{\rm pin}$, vortices are strongly pinned (the strongly pinned 
Bose glass phase) below the matching condition ($\rho = 0.25$) 
than above (the weakly pinned Bose glass phase). 
} 
\label{fig:6} 
\end{figure}

For high $V_{\rm pin}$ (say $V_{\rm pin}/t = 8.0$ in Fig.~\ref{fig:5}), 
we note that 
\begin{equation}
\label{eq:mott}
{\rho_s \over \rho} \Big\vert_{B > B_{\phi}}  > 
{\rho_s \over \rho} \Big\vert_{B < B_{\phi}}, 
\end{equation}
implying the vortices are more entangled 
above the matching condition, than below. 
Alternatively, in Fig.~\ref{fig:6} we plot binding energy per 
vortex defined as 
\begin{equation} 
B.E. = {E(V_{\rm pin}) - E(0) \over N_v} \  \cdot 
\label{eq:be} 
\end{equation} 

It is clear that above the matching field ($B_{\phi}$) 
vortices are less bound by disorder than below. 
So depending upon the induced $B$, compared with $B_{\phi}$, 
we find three phases within the BG, namely a strongly pinned
Bose glass (SBG for $B < B_{\phi}$), 
a Mott insulator (for $B = B_{\phi}$),  
and a weakly pinned Bose glass (WBG for $B > B_{\phi}$), 
consistent with ref.~\cite{leo}.   

In the  SBG phase, vortices are collectively pinned by columnar pins 
\cite{nelson-vinokur,leo}, whereas in the WBG phase, one has 
patches of ordered regions \cite{sen}. 
It is interesting to see that calculations on such small
systems, do exhibit signatures of these distinct glassy
phases. 

%
%

\subsection{Box Disorder} 
\protect{\label{subsec:box}} 

To further illustrate the effect of disorder on the FLL, 
we now present our results for the case where each site $i$ 
on the lattice has a random potential $\Delta_i$ chosen 
from a uniform distribution of $[- \Delta/2,\Delta/2]$.

\begin{figure}
\epsfxsize=\hsize \centerline{\epsfbox{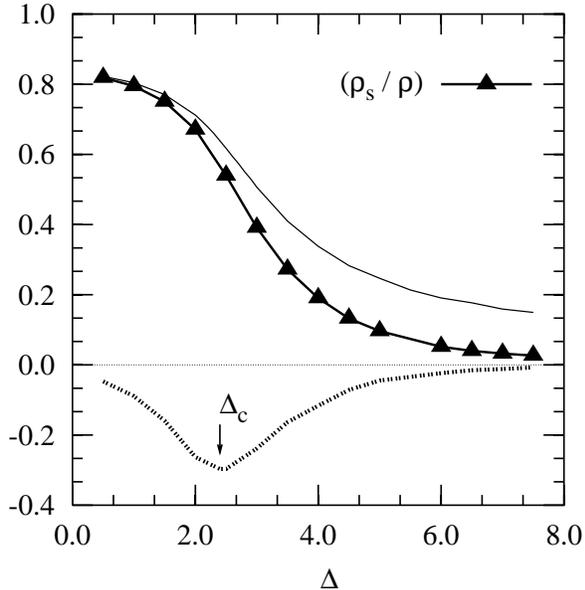}}
\caption{
{\bf Box Disorder}: 
 ``Superfluid" density $\rho_s / \rho \propto c_{44}^{-1}$
(solid triangles) as a function of $\Delta$ 
for $N_v=4; \rho=0.25$. 
Slope of $\rho_s (\Delta) / \rho$ curve is shown at the bottom (dotted line). 
The point of maximum slope is identified as $\Delta_c$ (shown by an arrow). 
The kinetic energy along the direction of the twist, 
$\langle \mbox{k.e.} \rangle_x / 2 t \rho$ (solid line), serves 
as an upper bound to $\rho_s / \rho$.  
Data averaged over 400 disorder realisations. 
}
\label{fig:2} 
\end{figure}

\begin{figure}
\epsfxsize=\hsize \centerline{\epsfbox{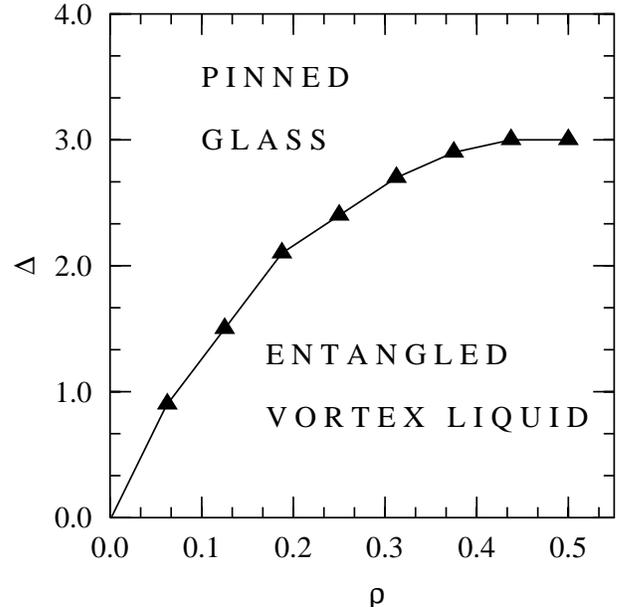}}
\caption{
{\bf Phase Diagram}: 
Locus of critical disorder $\Delta_c$, as a function of vortex density $\rho$. 
The line separates two phases: at low disorder, an entangled
``superfluid" phase of flux lines and at 
high disorder a pinned ``Bose-Glass" phase. 
The separatrix $\Delta_c (\rho)$ is determined by the emergence of a
finite fraction of incoherent sites with $\langle n_i \rangle = 0$ or $1$ 
Simulations are done on a $4 \times 4$ lattice, with periodic boundary 
conditions. Disorder ($\delta \mu_i$) is chosen from a uniform distribution
of width $[-\Delta/2, \Delta/2]$. 
Results are averaged over 400 disorder realizations. 
}
\label{fig:1} 
\end{figure}

In Fig.~\ref{fig:2} we plot the inverse tilt modulus 
$c_{44}^{-1}\propto \rho_s $
as a function 
of $\Delta$ for a fixed density of vortices ($\rho=0.25$). 
For small $\Delta$, $c_{44}$ has a finite
value (the entangled phase of vortices) and for large $\Delta$ 
it tends to diverge (the Bose glass phase of vortices).
In the thermodynamic limit, it is expected that 
$\rho_s \sim \vert \Delta - \Delta_c \vert^{\zeta}$ with 
$\zeta = \nu z $ in 2D, where $\nu$ is the correlation 
length exponent, and $z$ is the dynamical exponent \cite{fisher-weichmann}.
However, in a finite system, the transition is rounded.
As an indicator of the critical disorder strength $\Delta_c$, 
we choose the point of maximum slope of $\rho_s (\Delta) / \rho$
(shown on the  dotted curve at the bottom in Fig~\ref{fig:2}). 
A locus of $\Delta_c$ for different densities $\rho$ gives us
the phase boundary between the superfluid and the Bose glass
(see Fig.~\ref{fig:1}).

For zero disorder, $\rho_s / \rho$ differs from unity, which 
is a consequence of broken Galilean invariance. 
The kinetic energy per particle along the direction of the 
twist was shown to be an upper bound to $\rho_s / \rho$ \cite{arun}. 
We plot $\langle \rm{ke} \rangle_x / 2 t \rho$ as a function of 
$\Delta$ in the same figure. 
Near $\Delta = 0$ the kinetic energy per particle is
exactly equal to the superfluid density and with increasing disorder
it serves as an upper bound
but being a local quantity, it fails 
to pick up the transition\cite{arun}. 

%
%

\begin{figure}
\epsfxsize=\hsize \centerline{\epsfbox{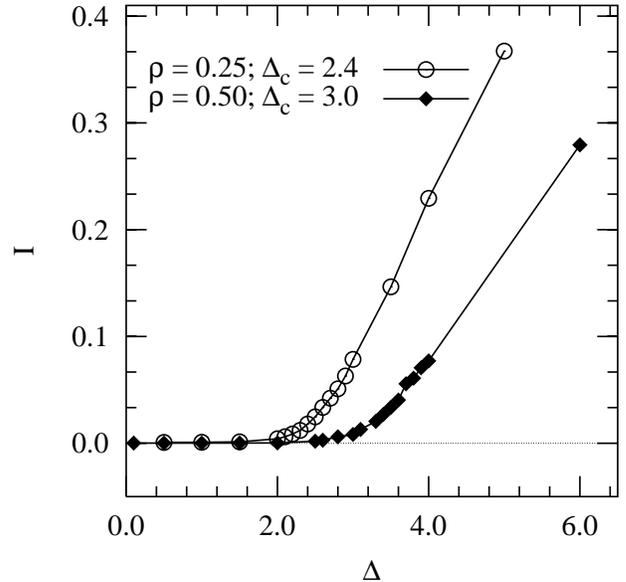}}
\caption{
{\bf Box Disorder}: 
Inverse Participation Ratio $I$ 
(see Eq.(\ref{eq:ipr}) for definition) which is a measure of 
the spatial extent of the boson wavefunction,  
for $N_v =4$ and $8$ as a function of $\Delta$. 
$I \sim 0$ for entangled vortices ($\Delta < \Delta_c$) and finite
for a pinned vortices ($\Delta > \Delta_c$).  
}
\label{fig:3} 
\end{figure}

A quantity of interest to probe the spatial extent of the wavefunction 
is the inverse participation ratio defined as 
\begin{equation} 
I = {\int \vert \psi_0 \vert^4 (r)  dr \over 
\left( \int \vert \psi_0 \vert^2 (r) dr \right)^2} \;,
\label{eq:ipr} 
\end{equation} 
where, $\psi_0 (r) $ is the ground state wavefunction 
of Eq.~(\ref{eq:Bosehubbard}).
For a localized state, $I$ has a finite value, whereas for an 
extended state, $I \sim 0$. 
In Fig.~\ref{fig:3} we plot $I$ as a function of $\Delta$ for 
$\rho = 0.25$ and 0.50. 
$I \sim 0$ till $\Delta_c$ and starts rising above $\Delta_c$ 
indicating localization or pinning of vortices. 
For high disorder, the states are localized for all densities. 
Though it should be noted that, such a numerical scheme {\em cannot} 
really distinguish between an extended state and a localized state with large
localization length (say for low $\Delta$). 

\begin{figure}
\epsfxsize=\hsize \centerline{\epsfbox{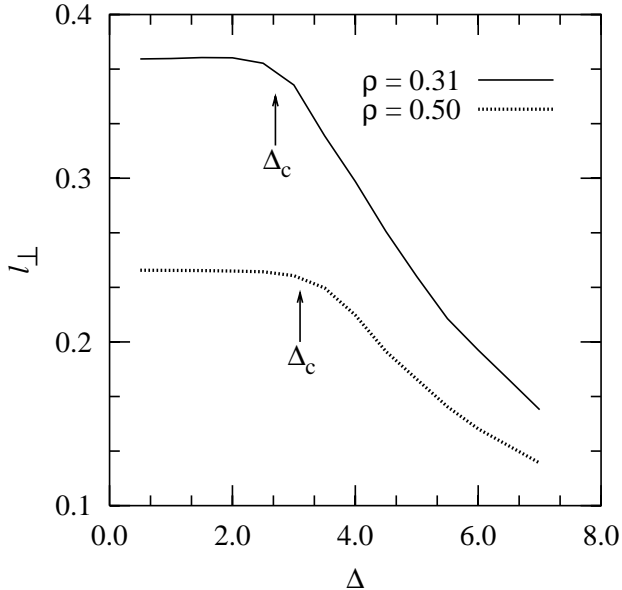}}
\caption{
{\bf Box Disorder}: 
Transverse wandering length per vortex $l_{\perp}$ 
(see Eq.~(\ref{eq:lperp1}) for definition) 
as a function of $\Delta$ for $N_v = 5$ and $8$. 
$l_{\perp}$ is a constant below $\Delta_c (\rho)$, almost equal to
the clean case, and reduces above $\Delta_c$ marking pinning of 
vortices. 
}
\label{fig:4} 
\end{figure}

The degree of wandering of a flux line transverse to its length
due to thermal fluctuations, 
can be obtained from the properties of the ground state 
wavefunction $\psi_0(r)$ in the mapped problem \cite{nelson-vinokur} 
by, 
\begin{equation} 
l_{\perp}^2 = \int d^2 r \; r^2 \psi_0^2 (r). 
\label{eq:lperp1} 
\end{equation} 
We plot the transverse wandering length per vortex in Fig.~\ref{fig:4} 
for 
$N_v = 5$ and $8$. 
Till $\Delta_c$, $l_{\perp}$ is constant (almost equal to the 
clean case). 
For $\Delta > \Delta_c$, $l_{\perp}$ reduces, thereby
marking localization of vortices due to disorder. 

%
%

The locus of critical disorder strength $\Delta_c$, 
estimated from the behaviour of quantities discussed above
as a function of $\Delta$, for various densities ($\rho$) 
gives us the phase
boundary between the superfluid (entangled liquid) and 
the Bose glass (pinned)
in the density-disorder ($\rho - \Delta$) plane shown 
in Fig.~\ref{fig:1}.

Another measure of the critical disorder strength 
$\Delta_c$ (for a given $\rho$)
is the disorder at which a finite fraction 
of {\it incoherent} sites occurs on the lattice.
In a superfluid, phase coherence is achieved due 
to tunneling of bosons between different sites leading to 
fractional occupancies on sites. 
The long wavelength excitations are gapless, but due
to the phase rigidity, local phase rotations cost a 
finite energy. 
We call a site {\em coherent} if the occupancy of that
site is fractional. 
On the other hand, at high $\Delta$, bosons are localised 
due to disorder. 
In this disorder dominated regime, local phase rotations
do not cost any energy at a site which is integer occupied. 
We call a site {\em incoherent} if $\langle n_i \rangle = 0$ or 1,
where $\langle n_i \rangle = 1$ corresponds to a single 
vortex pinned on the site $i$ all along its length. 

A locus of $\Delta_c (\rho)$ based on the emerging 
of an incoherent site concurs with the phase 
diagram shown in Fig.~\ref{fig:1}. 
The phase diagram, is re-entrant 
due to particle-hole symmetry about half filling ($\rho = 0.5$). 
At a mean field level, it has been shown that the system becomes
localized before clusters of nearly-incoherent sites percolate on the 
lattice \cite{shesh2}.

%
%

In this mean field theory, it is known that a Bose glass can 
never be accessed, because, a truly incoherent site with 
$\langle n_i \rangle =0$ can be obtained only at $\Delta = \infty$ 
\cite{shesh1}.
In ref.~\cite{arun}, by modeling the 
disordered boson problem as a network of random resistors,
it was argued that in order to stop the flow of the current 
(or destroy superfluidity), a line of resistors perpendicular to 
the direction of the flow must be cut. 
In this perspective, we note that, 
our criterion acts as a {\em precursor} to the 
actual SF-BG transition. 
The true SF-BG transition would lie above our $\Delta_c$ 
and below the percolation transition ($\Delta_c^p$) 
described in ref.~\cite{shesh2}.

The results presented above are for bounded disorder. 
For the case of unbounded (Gaussian) disorder, we find the 
same qualitative results as above.
In some cases, $\Delta_c$ is marginally reduced relative 
to the $\Delta_c$ for the bounded disorder, 
since one can get large disorder values from the tails of
the distribution. 

Now we discuss an interesting experimental implication of 
our phase diagram.
We see from Fig.~\ref{fig:1} that for a higher density of vortices 
a higher disorder strength is required
to prevent entanglement and wandering of the flux-lines. 
This happens because with increasing density ($\rho$), the disorder 
is effectively screened by vortices 
(strongest attractive sites are occupied first) and a new 
vortex introduced sees a weaker disorder 
landscape; thus is weakly pinned. 
Therefore, we conjecture that, 
{\em the transition temperature $T_{\rm BG}$, for a transition 
from a pinned Bose glass 
to an entangled liquid, will reduce with increasing
vortex density}.

%
%

\begin{figure}
\epsfxsize=\hsize \centerline{\epsfbox{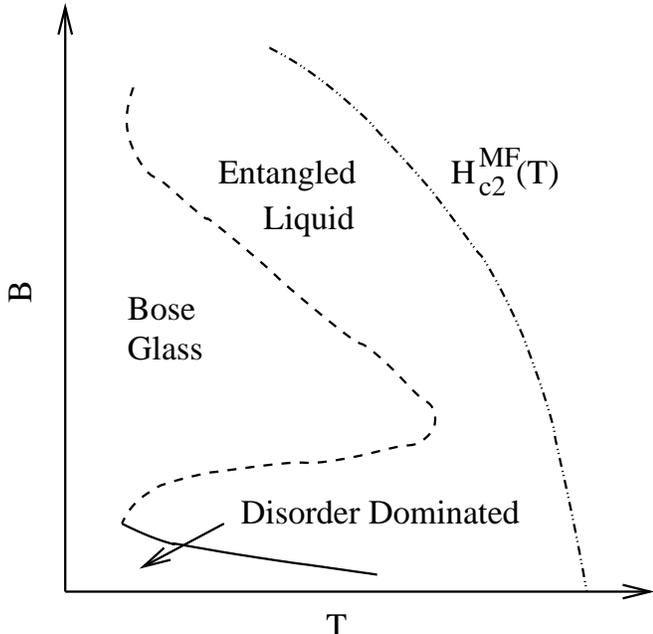}}
\caption{
Schematic Phase diagram for vortices in the $H-T$ plane. 
The flux line lattice melts much before $H_{c2}^{\rm MF}(T)$.
The melting line $T_{\rm BG}$ separates a pinned 
Bose-Glass and an entangled vortex liquid. 
The dashed line is the re-entrant melting line. 
Our conjecture for the $T_{\rm BG}$ is shown by the solid line,
in the disorder dominated, low $H$ region.  
The figure is not drawn to scale.}
\label{fig:7} 
\end{figure}

This conjecture is valid in two regimes in the $H-T$ plane. 
(a) At low fields, where the pinning energy 
dominates over interaction energy. 
For example, when the applied field is less than the 
matching field $B_{\phi}$, 
columnar pins outnumber vortices and the pinning energy 
($E_{\rm pin} \approx \epsilon_0 \ln (b / \xi)$, where 
$b$ is the diameter of the columnar pin and $\xi$ the 
coherence length) dominates over the magnetic repulsion \cite{leo}.  
We show this $T_{\rm BG}$ by a solid line in Fig.~\ref{fig:7}.
The transition line we conjecture has a similar shape to the 
Nelson-Le Doussal line \cite{nelson-ledoussal} {\em above} 
which point-defects play no role in deciding the structure of
vortices. 
The physical reasons for the origin of Nelson-Le Doussal
line and our $T_{\rm BG}$ are very different. 
Point defects ``promote" flux line wandering and 
entanglement, whereas correlated disorder ``inhibits" wandering
and promotes localization. 
And it is precisely this role of columnar pins which leads to 
the new $T_{\rm BG}$ at low fields. 
(b) At high fields, where vortex bundle pinning dominates over 
individual vortex pinning \cite{larkin-vinokur}. 
In our model, since we neglect the long-range part in 
the inter-vortex interaction, we do not 
see signatures of vortex bundle pinning at high densities. 
For the case of vortex-bundle pinning, the screening argument 
would still hold, and hence the topology of $T_{\rm BG}$ would be 
the same at high fields. 

At intermediate fields, the re-entrant ``nose" 
(shown by the dashed line in Fig.~\ref{fig:7})
was predicted by Nelson~\cite{nelson} and has been experimentally
verified ~\cite{ghosh}. 
We dont capture this re-entrant part in our simulations since 
we use an approximate form of the inter-vortex potential. 

To conclude, we conjecture a double re-entrant phase diagram.
Upon decreasing the field near the ``nose" the vortex liquid
localises into a glass where vortex bundles are pinned due
to disorder and then delocalizes again into an entangled liquid.
At lower fields, individual vortices are pinned by columnar 
defects, giving rise to another disorder dominated glassy phase.   

Finally we note that, the bosonic Hubbard model 
is a paradigm for studying quantum phase transitions \cite{tvr},
in systems such as $^4{\rm He}$ on random substrates, superconducting
disordered thin films \cite{haviland}, magnets and Josephson-junction
arrays \cite{mooij}.
The phase diagram in Fig.~\ref{fig:1}, for example is 
consistent with the 
field induced superconductor-insulator (SI) transition observed
in coupled Josephson-junction arrays \cite{mooij}. 
In low magnetic fields, vortices at $T=0$ are pinned by disorder
introduced via random offset charges.
With increasing fields, the vortex density increases and at some
critical density, vortices Bose-condense\cite{mpaf}. 
This vortex superfluid leads to an inifinite resistance, giving
way to an insulator. 

%
%
\section{Acknowledgments} 

We thank K. Sheshadri, Shobho Bhattacharya and Arun Grover for 
interesting discussions. 
AN acknowledges financial support from CSIR, India and would like to thank 
the Tata Institute of Fundamental Research, Mumbai for hospitality where 
part of this work was done. 
AN and DK also gratefully acknowledge partial financial support 
from Indo-French centre for the promotion
of Advanced Research (New Delhi)/Centre Franco Indian Pour la Promotion
de la Recherche Avanc\'ee.

%
%

\end{multicols}       
\end{document}